# Semantic Search by Latent Ontological Features

Tru H. Cao and Vuong M. Ngo

Faculty of Computer Science and Engineering
Ho Chi Minh City University of Technology
Viet Nam
{vuong.cs@gmail.com}

**Abstract.** Both named entities and keywords are important in defining the content of a text in which they occur. In particular, people often use named entities in information search. However, named entities have ontological features, namely, their aliases, classes, and identifiers, which are hidden from their textual appearance. We propose ontology-based extensions of the traditional Vector Space Model that explore different combinations of those latent ontological features with keywords for text retrieval. Our experiments on benchmark datasets show better search quality of the proposed models as compared to the purely keyword-based model, and their advantages for both text retrieval and representation of documents and queries.

## 1 Introduction

The usefulness and explosion of information on the WWW have been challenging research on information retrieval, regarding how that rich and huge resource of information should be exploited efficiently. Information retrieval is not a new area but still attracts much research effort, social and industrial interests, because, on the one hand, it is important for searching required information and, on the other hand, there are still many open problems to be solved to enhance search performance. Retrieval precision and recall could be improved by developing appropriate models, typically as similarity-based [9, 23], probabilistic relevance [27], or probabilistic inference [28] ones. Semantic annotation, representation, and processing of documents and queries are another way to obtain better search quality [6, 10, 12, 29].

Traditionally, text retrieval is only based on keywords (KW) occurring in documents and queries. Later on, word similarity and relationship are exploited to represent and match better documents to a query. Noticeably, words include those that represent named entities (NE), which are referred to by names such as people, organizations, and locations [24]. Together with keywords, co-occurring named entities in a text are indispensable part defining its content. In particular, in the top 10 search terms by YahooSearch[1] and GoogleSearch[2] in 2008, there are respectively 10 and 9 ones that are named entities.

In fact, keywords alone are not adequate, because named entities in a text and a query cover under their textual forms (i.e., names) ontological features that are significant to the semantics of the text and constitute the user intention in the query. Named entities and their properties are defined in an ontology and knowledge base[3] of

---

discourse. Firstly, it is the class of a named entity, for which texts containing "*Ha Noi*", "*Paris*", and "*Tokyo*" could be answers for a query about capital cities in the world. Searching purely based on keywords fails to do that because it does not use the common latent class information of such named entities to match with the class of named entities of user interest. Secondly, it is the identifier of a named entity, for which texts about "*U.S.*", "*USA*", "*United States*", and "*America*" should be returned for a query about the same country *United States of America*. Keyword-based searching also fails because it does not use the fact that an entity may exist under different aliases. If those latent ontological features of named entities, i.e., their classes, identifiers, and aliases, are annotated in texts then, for example, one can search for, and correctly obtain, web pages about *Washington* as a person. Whereas current search engines like Google may return any page that contains the word *Washington*, though it is the name of a state or a university.

As shown later in this paper, there are different combinations of ontological features of named entities that can be of user interest and expressed in a query. Nevertheless, usually, a query cannot be completely specified without keywords, like "*economic growth* of the *East Asian countries*", where *East Asian countries* represent named entities while *economic* and *growth* are keywords. It is thus natural and reasonable to combine named entities and keywords in representation of texts and queries to enhance search quality.

Until now, to our knowledge, there is no information retrieval model that formally takes into account all above-mentioned named entity features in combination with keywords. In this paper we propose ontology-based extensions of the Vector Space Model (VSM) that explore and analyse different combinations of ontological features and keywords. Implementation and experiments are also carried out to evaluate and compare the performance of the proposed models themselves and to the traditional purely keyword-based VSM.

Section 2 recalls the basic notion of the traditional VSM, and its extension to a multi-vector model for various named entity spaces. Section 3 presents alternative extended VSMs that combine both named entities and keywords. Section 4 is for evaluation and discussion on experimental results. In Section 5, we discuss related works in comparison with ours. Finally, Section 6 draws concluding remarks for the paper.

## 2    An Ontology-Based Multi-Vector Model

Despite having known disadvantages, VSM is still a popular model and a basis to develop other models for information retrieval, because it is simple, fast, and its ranking method is in general almost as good as a large variety of alternatives [1, 19]. We recall that, in the keyword-based VSM, each document is represented by a vector over a space of keywords of discourse. Conventionally, the weight corresponding to a term dimension of the vector is a function of the occurrence frequency of that term in the document, called *tf*, and the inverse occurrence frequency of the term across all the existing documents, called *idf*. The similarity degree between a document and a query is then defined as the cosine of their representing vectors.

With terms being keywords, the traditional VSM cannot satisfactorily represent the semantics of texts with respect to the named entities they contain, such as for the following queries:

$Q_1$: Search for documents about *Georgia*.

$Q_2$: Search for documents about *companies*.

$Q_3$: Search for documents about *locations named Washington*.

$Q_4$: Search for documents about *Moscow, Russia*.

Query $Q_1$ is to search for documents about any entity named *Georgia*, and correct answers include those about the state *Georgia* of the *USA* or the country *Georgia* next to *Russia*. However, documents about *Gruzia* are also relevant because *Gruzia* is another name of the country *Georgia*, which simple keyword-matching search engines miss. For query $Q_2$, a target document does not necessarily contain the keyword *company*, but only some named entities of the class *Company*, i.e., real commercial organizations in the world. For query $Q_3$, correct answers are documents about the state *Washington* or the capital *Washington* of the *USA*, which are locations, but not those about people like President *Washington*. Meanwhile, query $Q_4$ targets at documents about a precisely identified named entity, i.e., the capital *Moscow* of *Russia*, not other cities also named *Moscow* elsewhere.

For formally representing documents (and queries) by named entity features, we define the triple ($N$, $C$, $I$) where $N$, $C$, and $I$ are respectively the sets of names, classes, and identifiers of named entities in an ontology of discourse. Then:

1. Each document $d$ is modelled as a subset of $(N \cup \{*\}) \times (C \cup \{*\}) \times (I \cup \{*\})$, where '*' denotes an unspecified name, class, or identifier of a named entity in $d$, and

2. $d$ is represented by the quadruple ($\vec{d}_N$, $\vec{d}_C$, $\vec{d}_{NC}$, $\vec{d}_I$), where $\vec{d}_N$, $\vec{d}_C$, $\vec{d}_{NC}$, and $\vec{d}_I$ are respectively vectors over $N$, $C$, $N \times C$, and $I$.

A feature of a named entity could be unspecified due to the user intention expressed in a query, the incomplete information about that named entity in a document, or the inability of an employed NE recognition engine to fully recognize it [22]. Each of the four component vectors introduced above for a document can be defined as a vector in the traditional *tf.idf* model on the corresponding space of entity names, classes, name-class pairs, or identifiers, instead of keywords. However, there are two following important differences with those ontological features of named entities in calculation of their frequencies:

1. The frequency of a name also counts identical entity aliases. That is, if a document contains an entity having an alias identical to that name, then it is assumed as if the name occurred in the document. For example, if a document refers to the country *Georgia*, then each occurrence of that entity in the document is counted as one occurrence of the name *Gruzia*, because it is an alias of *Georgia*. Named entity aliases are specified in a knowledge base of discourse.

2. The frequency of a class also counts occurrences of its subclasses. That is, if a document contains an entity whose class is a subclass of that class, then it is assumed as if the class occurred in the document. For example, if a document refers to *Washington DC*, then each occurrence of that entity in the document is counted as one occurrence of the class *Location*, because *City* is a subclass of *Location*. The class subsumption is defined by the class hierarchy of an ontology of discourse.

The similarity degree of a document $d$ and a query $q$ is then defined to be:

$$sim(\vec{d}, \vec{q}) = w_N.cosine(\vec{d}_N, \vec{q}_N) + w_C.cosine(\vec{d}_C, \vec{q}_C) + w_{NC}.cosine(\vec{d}_{NC}, \vec{q}_{NC}) +$$

$$w_I.cosine(\vec{d}_I, \vec{q}_I) \qquad \text{(Eq. 1)}$$

where $w_N + w_C + w_{NC} + w_I = 1$. We deliberately leave the weights in the sum unspecified, to be flexibly adjusted in applications, depending on developer-defined relative significances of the four ontological features.

We note that the join of $\vec{d}_N$ and $\vec{d}_C$ cannot replace $\vec{d}_{NC}$ because the latter is concerned with entities of certain name-class pairs (e.g. the co-occurrence of an entity named *Georgia* and another country mention in a text does not necessarily refer to the country *Georgia*). Meanwhile, $\vec{d}_{NC}$ cannot replace $\vec{d}_I$ because there may be different entities of the same name and class (e.g. there are different cities named *Moscow* in the world). Also, since names and classes of an entity are derivable from its identifier, products of $I$ with $N$ or $C$ are not included. In brief, here we generalize the notion of terms being keywords in the traditional VSM to be entity names, classes, name-class pairs, or identifiers, and use four vectors on those spaces to represent a document or a query for text retrieval. Figure 2.1 shows a query in the TIME test collection (available with SMART [3]) and its corresponding sets of ontological terms, where *InternationalOrganization_T.17* is the identifier of *United Nations* in a knowledge base of discourse.

---

Query: "*Countries have newly joined the United Nations*".

Ontological term set:
{(*/*Country*/*), (*/*/*InternationalOrganization_T.17*)}

---

**Fig. 2.1.** Ontological terms extracted from a query

## 3 Combining Named Entities and Keywords

Clearly, named entities alone are not adequate to represent a text. For example, in the query in Figure 2.1, *joined* is a keyword to be taken into account, and so are *Countries* and *United Nations*, which can be concurrently treated as both keywords and named entities. Therefore, a document can be represented by one vector on keywords and four vectors on ontological terms. The similarity degree of a document $d$ and a query $q$ is then defined as follows:

$$sim(\vec{d}, \vec{q}) = \alpha.[w_N.cosine(\vec{d}_N, \vec{q}_N) + w_C.cosine(\vec{d}_C, \vec{q}_C) + w_{NC}.cosine(\vec{d}_{NC}, \vec{q}_{NC}) +$$

$$w_I.cosine(\vec{d}_I, \vec{q}_I)] + (1-\alpha).cosine(\vec{d}_{KW}, \vec{q}_{KW}) \qquad \text{(Eq. 2)}$$

where $w_N + w_C + w_{NC} + w_I = 1$, $\alpha \in [0, 1]$, and $\vec{d}_{KW}$ and $\vec{q}_{KW}$ are respectively the vectors representing the keyword features of $d$ and $q$. The coefficient $\alpha$ weights relative importance of the NE and KW components in document and query representations. We denote this multi-vector model combining named entities and keywords by KW$\cup$NE.

Furthermore, we explore another extended VSM that combines keywords and named entities. That is we unify and treat all of them as *generalized terms*, where a term is counted either as a keyword or a named entity but not both. Each document is then represented by a single vector over that generalized term space. Document vector representation, filtering, and ranking are performed as in the traditional VSM, except for taking into account entity aliases and class subsumption as presented in Section 2. We denote this model by KW+NE. Figure 3.1 show another query in the TIME test collection and its corresponding key term sets for the multi-vector model and the generalized term model.

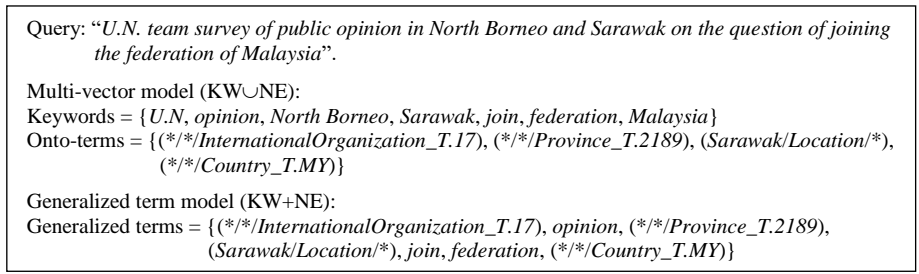

Query: "*U.N. team survey of public opinion in North Borneo and Sarawak on the question of joining the federation of Malaysia*".

Multi-vector model (KW∪NE):
Keywords = {*U.N, opinion, North Borneo, Sarawak, join, federation, Malaysia*}
Onto-terms = {(*\*/\*/InternationalOrganization_T.17*), (*\*/\*/Province_T.2189*), (*Sarawak/Location/\**), (*\*/\*/Country_T.MY*)}

Generalized term model (KW+NE):
Generalized terms = {(*\*/\*/InternationalOrganization_T.17*), *opinion*, (*\*/\*/Province_T.2189*), (*Sarawak/Location/\**), *join, federation*, (*\*/\*/Country_T.MY*)}

**Fig. 3.1.** Keywords, ontological terms, and generalized terms extracted from a query

The system architecture of NE-based text retrieval is shown in Figure 3.2. It contains an ontology and knowledge base of named entities in a world of discourse. The NE Recognition and Annotation module extracts and embeds information about named entities in a raw text, before it is indexed and stored in the NE-Annotated Text Repository. Users can search for documents about named entities of interest via the NE-Based Text Retrieval module.

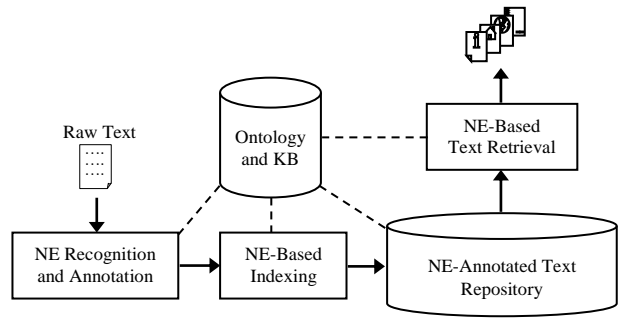

**Fig. 3.2.** System architecture for NE-based text retrieval

We have implemented the above-extended VSMs by employing and modifying Lucene, which is a general open source for storing, indexing and searching documents [11]. In fact, Lucene uses the traditional VSM with a tweak on the document magnitude term in the cosine similarity formula for a query and a document. In Lucene, a term is a character string and term occurrence frequency is computed by exact string matching, after keyword stemming and stop-word removal. Here are our modifications of Lucene for what we call *S-Lucene* for the above-extended VSMs:

1. Indexing documents over the four ontological spaces corresponding to *N*, *C*, *N×C*, and *I*, and the generalized term space, besides the ordinary keyword space, to support the new models.
2. Modifying Lucene codes to compute dimensional weights for the vectors representing a document or a query, in accordance to each of the new models.
3. Modifying Lucene codes to compute the similarity degree between a document and a query, in accordance to each of the new models.

Each document is automatically processed, annotated, and indexed as follows:
1. Stop-words in the document are removed using a built-in function in Lucene.
2. The document is annotated with the named entities recognized by an employed NE recognition engine. For the multi-vector model, recognized entity names are also counted as keywords, but not for the generalized term model.
3. Taking into account entity aliases and class subsumption, a document is extended with respect to each entity named *n* possibly with class *c* and identifier *id* in the document as follows:
   - For the multi-vector model, the values *n*, *c*, (*n*, *c*), *alias*(*n*), *super*(*c*), (*n*, *super*(*c*)), (*alias*(*n*), *c*), (*alias*(*n*), *super*(*c*)), and *id* are added for the document.
   - For the generalized term model, the triples (*n*/*/*), (*/c*/*), (*n*/c*/*), (*alias*(*n*)/*/*), (*/super*(*c*)/*), (*n*/super*(*c*)/*), (*alias*(*n*)/c*/*), (*alias*(*n*)/ super*(*c*)/*), and (*/*/id*) are added for the document.
4. The extracted keywords, named entity values and triples in the document are indexed using the newly developed functions in S-Lucene.

Here *alias*(*n*) and *super*(*c*) respectively denote any alias of *n* and any super class of *c* in an ontology and knowledge base of discourse. For *super*(*c*), we exclude the top-level classes, e.g. *Entity*, *Object*, *Happening*, *Abstract*, because they are too general and could match many named entities.

For each query, after stop-word removal and NE recognition and annotation, it is processed further by the following steps:
1. Each recognized entity named *n* possible with class *c* and identifier *id* is represented by one or more named entity triples as follows:
   - For the multi-vector model, the most specific named entity annotation is used. We note that *id* is more specific than (*n*, *c*), which is more specific than both *c* and *n*.
   - For the generalized term model, the most specific and available triple among (*n*/*/*), (*/c*/*), (*n*/c*/*), and (*/*/id*) is used for the query.
2. The interrogative word *Who*, *What*, *Which*, *When*, *Where*, or *How*, if exists in the query, is mapped to an unspecified named entity of an appropriate class, as explained in details in the experimentation with the TREC dataset next.

## 4 Experimentation

### 4.1 Performance measures

We have evaluated and compared the new models in terms of the precision-recall (P-R) curve, F-measure-recall (F-R) curve, and single mean average precision (MAP).

For each query in a test collection, we adopt the common method in [18] to obtain the corresponding P-R and F-R curves. Meanwhile, MAP is a single measure of retrieval quality across recall levels and considered as a standard measure in the TREC community [30].

In order to confirm that compared systems with different observed quality measures actually have different performances, a statistical significance test is required [13]. In [26], the authors compared the five significance tests that have been used by researchers in information retrieval, namely, Student's paired t-test, Wilcoxon signed rank test, sign test, bootstrap, and Fisher's randomization (permutation). They recommended Fisher's randomization test for evaluating the significance of the observed difference between two systems. As shown therein, 100,000 permutations were acceptable for a randomization test and the threshold 0.05 could detect significance. In this work, we adopt that test for pairs of systems under consideration.

### 4.2 Testing with the TIME dataset

We have first evaluated the proposed models on the TIME collection [3], with 83 queries and 425 documents. The ontology, knowledge base, and NE recognition engine of KIM [17] are employed to automatically annotate named entities in documents. The ontology consists of 250 classes and the knowledge base contains about 40,000 instances. The average precision and recall of the NE recognition engine are about 90% and 86%, respectively[4]. In the experiments, we set the weights $w_N = w_C = w_{NC} = w_I = 0.25$ and $\alpha = 0.5$, assuming that the keyword and named entity dimensions are of equal importance. Almost all queries (80 out of 83) in this test dataset do not contain interrogative words, so we do not apply mapping interrogative words to named entity classes in this test.

Table 4.1 and Table 4.2 respectively present the average precision values and average F-measure values of the purely keyword-based VSM (KW), also implemented in the same Lucene, the multi-vector model, and the generalized term model, at each of the standard recall levels. Table 4.3 shows their MAP values on which one can have the following observations.

Firstly, the purely NE-based model and the purely keyword-based model have little different MAP values, which are significantly lower than those of the combined keyword and named entity models. Secondly, the KW+NE model has the highest MAP value. We are interested in KW+NE not only because of its MAP value, but also because of its simplicity and uniformity as compared to the multi-vector model. So we have further conducted a randomization test for KW+NE against every other model. In Table 4.4, $|MAP(A) - MAP(B)| = \delta$ is the observed difference between two models $A$ and $B$; $N^-$ and $N^+$ are respectively the numbers of measured differences, out of 100,000 permutations, that are less than or equal to $-\delta$ and greater than or equal to $\delta$.

---

[4] It is reported at http://www.ontotext.com/kim/performance.html.

**Table 4.1.** The average precisions at the eleven standard recall levels on the TIME dataset

| Model | Recall (%) | | | | | | | | | | | |
|---|---|---|---|---|---|---|---|---|---|---|---|---|
| | 0 | 10 | 20 | 30 | 40 | 50 | 60 | 70 | 80 | 90 | 100 | |
| KW | 74.04 | 74.04 | 73.44 | 70.85 | 68.78 | 65.79 | 58.42 | 55.02 | 53.3 | 50.88 | 49.88 | Precision (%) |
| NE | 71.03 | 69.72 | 68.95 | 66.11 | 64.72 | 63.66 | 59.41 | 56.56 | 54.73 | 53.06 | 52.87 | |
| KW∪NE | 81.23 | 81.01 | 80.09 | 76.85 | 76.23 | 74.93 | 68.8 | 63.52 | 60.95 | 59.28 | 58.88 | |
| KW+NE | 82.67 | 82.3 | 80.98 | 78.84 | 77.02 | 75.19 | 71.37 | 69.13 | 67.03 | 63.89 | 63.18 | |

**Table 4.2.** The average F-measures at the eleven standard recall levels on the TIME dataset

| Model | Recall (%) | | | | | | | | | | | |
|---|---|---|---|---|---|---|---|---|---|---|---|---|
| | 0 | 10 | 20 | 30 | 40 | 50 | 60 | 70 | 80 | 90 | 100 | |
| KW | 0 | 16.2 | 28.52 | 37.85 | 45.36 | 50.66 | 51.82 | 53.62 | 55.88 | 56.65 | 58.2 | F-measure (%) |
| NE | 0 | 16.2 | 28.23 | 36.95 | 44.18 | 49.88 | 52.37 | 54.79 | 57.07 | 58.45 | 60.86 | |
| KW∪NE | 0 | 16.96 | 30.12 | 40.18 | 48.71 | 55.61 | 58.14 | 59.8 | 62.01 | 63.88 | 66.51 | |
| KW+NE | 0 | 16.93 | 30.18 | 40.66 | 48.96 | 55.36 | 59.46 | 63.1 | 66.17 | 67.4 | 70.06 | |

**Table 4.3.** The mean average precisions on the TIME dataset

| Model | KW | NE | KW∪NE | KW+NE |
|---|---|---|---|---|
| **MAP** | 0.6167 | 0.6039 | 0.6977 | 0.7252 |

**Table 4.4.** Two-sided p-values of randomization tests between the KW+NE model and the others on the TIME dataset

| Model $A$ | Model $B$ | $|\mathbf{MAP}(A) - \mathbf{MAP}(B)|$ | $\mathbf{N}^-$ | $\mathbf{N}^+$ | Two-Sided P-Value |
|---|---|---|---|---|---|
| KW+NE | KW | 0.1085 | 0 | 5 | 0.00005 |
| | NE | 0.1213 | 1 | 12 | 0.00013 |
| | KW∪NE | 0.0275 | 7,977 | 25,059 | 0.33036 |

Given the significance level threshold of 0.05, the results show that the KW+NE model truly performs better than the KW and NE models, while the differences of its MAP value from that of the KW∪NE model might be due to noises in the experiment. Figure 4.1 illustrates the average P-R and average F-R curves of the KW, NE, and KW+NE models. Figure 4.2 shows the per query differences in average precision between KW+NE and the other two models, where each dot above the horizontal axis corresponds to a query for which KW+NE performs better.

We note that the performance of any system relying on named entities to solve a particular problem partly depends on that of the NE recognition module in a preceding stage. However, in research for models or methods, the two problems should be separated. This paper is not about NE recognition and our experiments incur errors of the employed KIM ontology and annotation engine. Also, the focus of our work is on how relatively better a basic model enhanced with named entities is in comparison to the purely keyword-based one. In this paper, we choose the popular VSM as such a basic model, but other models could be used as alternative ones.

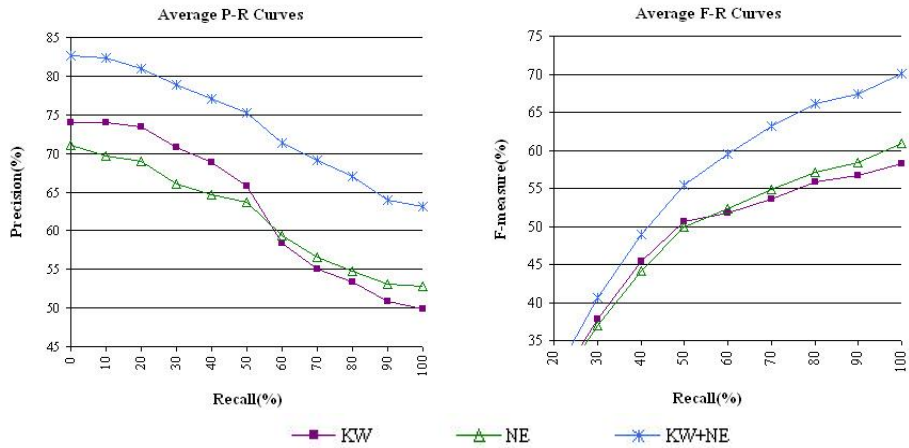

**Fig. 4.1.** Average P-R and F-R curves of KW, NE, and KW+NE on the TIME dataset

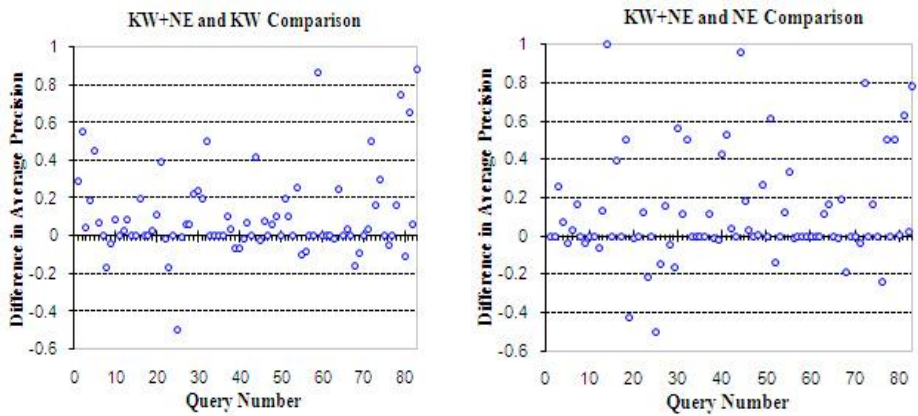

**Fig 4.2.** The per query differences in average precision of KW+NE with KW and NE

The generalized term model KW+NE is straightforward and simple, unifying keywords and named entities as generalized terms. Meanwhile, the multi-vector model KW∪NE with a comparable performance can be useful for clustering documents into a hierarchy via top-down phases each of which uses one of the four NE-based vectors presented above (cf. [4]).

For example, given a set of geographical documents, one can first cluster them into groups of documents about rivers and mountains, i.e., clustering with respect to entity classes. Then, the documents in the river group can be clustered further into subgroups each of which is about a particular river, i.e., clustering with respect to entity identifiers. As another example of combination of clustering objectives, one can first make a group of documents about entities named *Saigon*, by clustering them with respect to entity names. Then, the documents within this group can be clustered further into subgroups for *Saigon City*, *Saigon River*, and *Saigon Market*, for instance, by clustering them with respect to entity classes.

Another advantage of splitting document representation into four component vectors is that, searching and matching need to be performed only on those components that are relevant to a certain query. For example, in searching for documents about country capitals in the world, i.e., only entity classes are of concern, only the component vector $\vec{d}_C$ of a document $d$ should be used for matching with the corresponding class vector of the query.

### 4.3 Testing with a TREC dataset

We have then tested the generalized term model KW+NE on a larger dataset and on queries containing interrogative words, which also cover ontological information significant to retrieval. We have chosen the TREC L.A. Times document collection, consisting of more than 130,000 documents in nearly 500MB, and 124 queries out of 200 queries in the TREC QA Track 1999 that have answers in that document collection. We use KW+NE+Wh to denote the KW+NE model enhanced with mapping interrogative words to entity classes. KIM ontology, knowledge base, and NE recognition engine are also employed for this experiment.

**Table 4.5.** Mapping interrogative words to entity classes

| Interrogative Word | NE Class | Example Query |
|---|---|---|
| Who | Person | Who is the author of the book, "The Iron Lady: A Biography of Margaret Thatcher"? |
| | Woman | Who was the lead actress in the movie "Sleepless in Seattle"? |
| Which | Person | Which former Ku Klux Klan member won an elected office in the U.S.? |
| | City | Which city has the oldest relationship as a sister-city with Los Angeles? |
| Where | Location | Where did the Battle of the Bulge take place? |
| | WaterRegion | Where is it planned to berth the merchant ship, Lane Victory, which Merchant Marine veterans are converting into a floating museum? |
| When | DayTime | When did the Jurassic Period end? |
| | CalendarMonth | When did Beethoven die? |
| What | CountryCapital | What is the capital of Congo? |
| | Percent | What is the legal blood alcohol limit for the state of California? |
| | Money | What was the monetary value of the Nobel Peace Prize in 1989? |
| | Person | What two researchers discovered the double-helix structure of DNA in 1953? |
| How | Money | How much could you rent a Volkswagen bug for in 1966? |

**Original text**

Query: "*Who is the president of Stanford University?*"

Document: "*The California Compact ... and has been in existence for several years. The California group is co-chaired by Stanford University President Don Kennedy and ...*"

**KW keyword sets**

Query = {*president, Stanford, University*}

Document = {*California, Compact, existence, year, group, co-chair, Stanford, University, President, Don, Kennedy*}

**Annotated named entities**

1. *Who* presents a named entity of the class *Person*.
2. *Stanford University* is a named entity represented by (*Stanford University/University/ University_T.52*) and its alias (*Stanford/University/ University_T.52*).
3. *California Compact* is an un-identified named entity represented by (*California Compact/Organization/\**).
4. *California* is a named entity represented by (*California/Provincel Province_T.4198*).
5. *Don Kennedy* is an un-identified named entity represented by (*Don Kennedy/Man/\**).

**Part of the ontology**

1. *Person* is a class.
2. *University* has super-classes *EducationalOrganization*, *Organization*, *Group*, *Agent*.
3. *Organization* has super-classes *Group*, *Agent*.
4. *Province* has super-classes *PoliticalRegion*, *Location*.
5. *Man* has super-classes *Person*, *Agent*.

**KW+NE generalized term sets in Query**

Query = {*president* OR (*\*/\*/University_T.52*)}

**KW+NE+Wh generalized term sets in Query**

Query = {(*\*/Person/\**) OR *president* OR (*\*/\*/University_T.52*)}

**KW+NE and KW+NE+Wh generalized term sets in Document**

Document = {*existence, year, group, co-chair, President*}

∪{(*California Compact/\*/\**), (*\*/Organization/\**), (*California Compact/Organization/\**), (*\*/Group/\**), (*\*/Agent/\**), (*California Compact/Group/\**), (*California Compact/Agent/\**)}

∪{(*\*/Province_T.4198*), (*California/\*I\**), (*\*/Provincel\**), (*\*/PoliticalRegion/\**), (*\*/Location/\**), (*California/PoliticalRegion/\**), (*California/ Location/ \**)}

∪{(*\*/\*/University_T.52*), (*Stanford University/\*/\**), (*\*/University/\**), (*Stanford University/University/\**), (*Stanford/\*/\**), (*\*/EducationalOrganization/\**), (*\*/Organization/\**), (*\*/Group/\**), (*\*/Agent/\**), (*Stanford University/ EducationalOrganization/\**), (*Stanford University/ Organization/\**), (*Stanford University/ Group/\**), (*Stanford University/Agent/\**), (*Stanford/University/\**), (*Stanford/EducationalOrganizatio /\**), (*Stanford/Organization/\**), (*Stanford/Group/\**), (*Stanford/Agent/\**)}

∪{(*Don Kennedy/Man/\**), (*Don Kennedy/\*/\**), (*\*/Man/\**), (*\*/Person/\**), (*Don Kennedy/ Person/\**)}

**Fig. 4.3.** Keywords and named entity terms extracted from a query and a document

Most of those queries (119 out of 124) are in the form of questions by the interrogative words *Who*, *What*, *Which*, *When*, *Where*, and *How*. They actually represent the named entities of certain classes in question and thus are significant regarding whether a text contains them or not. Table 4.5 gives some examples on mapping interrogative words to entity classes, which are dependent on a query context. In the scope of this paper, for the experiments we manually map those words to certain entity classes, but it could be automatically done with high accuracy using the method proposed in [5]. Figure 4.3 shows the keyword sets for the KW model, the KW+NE generalized term sets, and the KW+NE+Wh generalized term sets of a query and part of a document in the test dataset, with annotated named entities in conjunction to part of the ontology of discourse.

Table 4.6 presents, and Figure 4.4 plots, the average precisions and F-measures of the KW, KW+NE, and KW+NE+Wh models at each of the standard recall levels. Figure 4.5 shows the per query differences in average precision between KW+NE+Wh and the other two models. The MAP values of the models in Table 4.7 and the two-sided p-values of randomization tests between them in Table 4.8 (with 100,000 permutations of two compared systems and the significance level threshold of 0.05) show that taking into account latent ontological features in queries and documents does enhance text retrieval performance; KW+NE+Wh performs about 10.8% better than KW in terms of MAP values.

The small difference between the MAP values of KW+NE+Wh and KW+NE (about 3.35%) could be due to the following facts. First, only 68 out of the 124 test queries have interrogative words mapped to NE classes, because 5 queries do not have interrogative words and 51 queries do not have corresponding NE classes in the employed ontology for their interrogative words. Second, for those 68 queries, KW+NE+Wh is better than, as good as, or worse than KW+NE in 32, 24, and 12 queries, respectively.

**Table 4.6.** The average precisions and F-measures at the eleven standard recall levels on the TREC dataset

| Measure | Model | Recall (%) | | | | | | | | | | |
|---------|-------|------|------|------|------|------|------|------|------|------|------|------|
| | | 0 | 10 | 20 | 30 | 40 | 50 | 60 | 70 | 80 | 90 | 100 |
| **Precision** (%) | KW | 66.3 | 66.2 | 63.6 | 60.1 | 56.5 | 54.9 | 45.7 | 40.4 | 38.0 | 37.4 | 36.9 |
| | KW+NE | 68.9 | 68.9 | 66.6 | 63.4 | 60.2 | 58.0 | 49.5 | 45.2 | 43.4 | 42.8 | 42.0 |
| | KW+NE+Wh | 72.2 | 71.9 | 69.5 | 65.3 | 62.0 | 60.6 | 52.3 | 47.9 | 46.1 | 45.0 | 44.3 |
| **F-measure** (%) | KW | 0.0 | 15.5 | 26.7 | 34.8 | 40.1 | 45.0 | 43.4 | 42.1 | 41.8 | 43.0 | 44.1 |
| | KW+NE | 0.0 | 16.0 | 27.6 | 36.0 | 41.6 | 46.4 | 45.7 | 45.3 | 46.0 | 47.6 | 48.4 |
| | KW+NE+Wh | 0.0 | 16.3 | 28.4 | 37.1 | 42.8 | 48.3 | 48.0 | 47.7 | 48.5 | 49.8 | 50.8 |

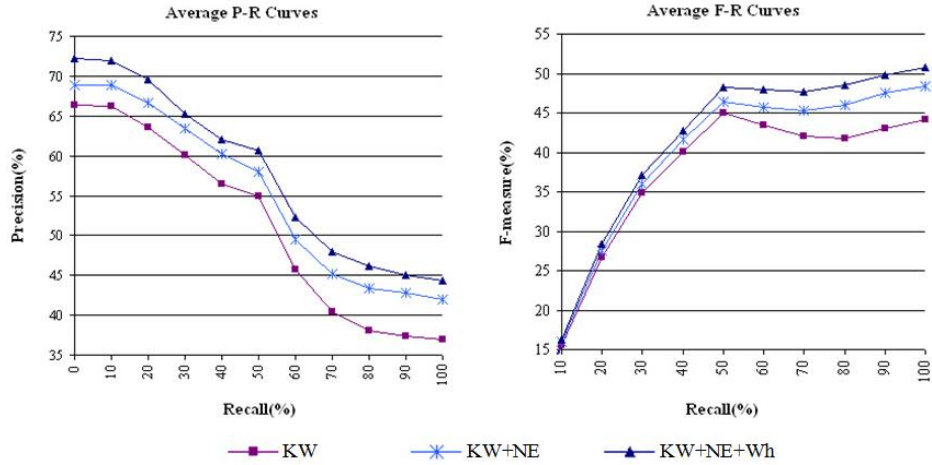

**Fig 4.4.** Average P-R and F-R curves of KW, KW+NE, and KW+NE+Wh on the TREC dataset

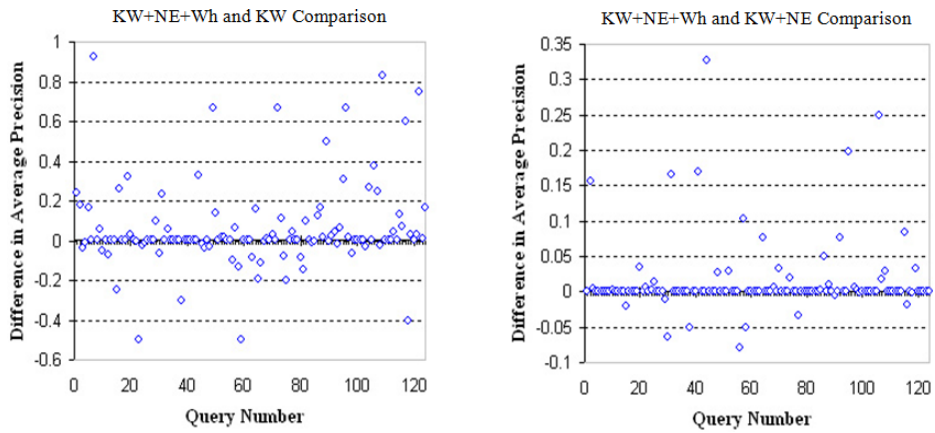

**Fig 4.5** The per query differences in average precision of KW+NE+Wh with KW and KW+NE

**Table 4.7.** The mean average precisions on the TREC dataset

| Model | KW | KW+NE | KW+NE+Wh |
|-------|-----|-------|----------|
| **MAP** | 0.50991 | 0.54691 | 0.5652 |

**Table 4.8.** Two-sided p-values of randomization tests between KW+NE+Wh, KW+NE, and KW on the TREC dataset

| Model *A* | Model *B* | \|**MAP(*A*)** − **MAP(*B*)**\| | **N⁻** | **N⁺** | **Two-Sided P-Value** |
|-----------|-----------|-------------------------------|--------|--------|------------------------|
| KW+NE+Wh | KW+NE | 0.0183 | 77 | 52 | 0.00129 |
| KW+NE+Wh | KW | 0.0553 | 143 | 259 | 0.00402 |
| KW+NE | KW | 0.037 | 1751 | 2500 | 0.04251 |

# 5    Related Works

In [31], each concept in a text was linked to its candidate concepts in Wikipedia and the text representation was enriched by the synonyms, hyponyms and associative concepts of those candidate concepts. For concepts representing named entities, the use of synonyms and hyponyms is similar to the use of aliases and super-classes for text extension in our NE-based models. In [14], the authors proposed a knowledge-based vector space model that took into account semantic similarities between terms in documents. These works are however for document clustering, but not document retrieval.

In the domain of geographic information systems, [15] briefly reported experiments to examine the impact of geographic features, in particular place names, on web retrieval performance. Also using named entity features, the Falcons system described in [7] was not for document retrieval, but provided a friendly interface for users to specify the properties of the objects to be searched for.

In [20], a probabilistic relevance model was introduced for searching passages about certain biomedical entity types (i.e., classes) only, such as genes, diseases, or drugs. Also in the biomedical domain, the similarity-based model in [32] considered concepts being genes and medical subject headings, such as *purification*, *HNF4*, or *hepatitis B virus*. Concept synonyms, hypernyms, and hyponyms were taken into account, which respectively corresponded to entity aliases, super-classes, and subclasses in our NE-based models. A document or query was represented by two component vectors, one of which was for concepts and the other for words. A document was defined as being more similar to a query than another document if the concept component of the former was closer to that of the query. If the two concept components were equally similar to that of the query, then the similarity between the word components of the two documents and that of the query would decide. However, as such, the word component was treated as only secondary in the model, and its domain was just limited within biomedicine.

Recently, [2] developed a search engine based on keywords and named entities of specified classes in texts. It appears that the work considered only entity classes in combination with keywords. Also, it did not present an underlying model like ours regarding how queries and documents are represented and their similarity computed. Moreover, it was about search efficiency rather than search quality, as only simple queries comprising a few keywords and entity classes were used for testing the precision and recall of the engine. In [8], the targeted problem was to search for named entities of specified classes associated with keywords in a query. For example, the given query "*Amazon Customer Service Phone*" therein, where *Phone* represented a named entity in question of the class *PhoneNumber*, was to search for the right phone number of *Amazon Customer Service* in web pages, while there could be other phone numbers in the same web pages too. As such, in contrast to ours, that work considered only entity classes and was not about searching for documents whose contents match a query. Meanwhile, [16] researched and showed that NE normalization improved retrieval performance. The work however considered only entity names and that normalization issue was in fact what we call name aliasing here.

Closely related works to ours are [21], [6] and [10]. As an early proposal, [21] enriched queries and texts with NE tags, which were used together with usual

keywords for text retrieval. Interrogative words were also replaced by corresponding NE tags. However, NE tags used therein were simply NE classes only. Also, class subsumption and name aliasing were not considered as in our work here. Experimental results showed that such NE tagging enhanced the relevance of documents retrieved. Nevertheless, the work used a variation of the precision measure, which was defined to be 0 if no relevant documents were found, or $1/N$ otherwise, with $N$ being the number of documents retrieved. Therefore its model is subsumed by, and its performance figures are not comparable to, ours.

In [6], the authors adapted the traditional VSM with vectors over the space of NE identifiers in a knowledge base of discourse. For each document or query, the authors also applied a linear combination of its NE-identifier-based vector and keyword-based vector with the equal weights of 0.5. The system was tested on the authors' own dataset. The main drawback was that every query had to be posed using RDQL, a query language for RDF, to first look up in the system's knowledge base those named entities that satisfied the query, before its vector could be constructed. For example, given the query searching for documents about *Basketball Player*, its vector would be defined by the basketball players identified in the knowledge base. This step of retrieving NE identifiers was unnecessarily time consuming. Moreover, a knowledge base is usually incomplete, so documents containing certain basketball players not existing in the knowledge base would not be returned. In our proposed models, the query and document vectors on the entity class *Basketball Player* can be constructed and matched right away.

Meanwhile, the LRD (Latent Relation Discovery) model proposed in [10] used both keywords and named entities as terms for a single vector space. The essential of the model was that it enhanced the content description of a document by those terms that did not exist, but were related to existing terms, in the document. The relation strength between terms was based on their co-occurrence. The authors tested the model on 20 randomly chosen queries from 112 queries of the CISI dataset [3] with 1460 documents selected from [25], achieving the maximum F-measure of 19.3. That low value might be due to the dataset containing few named entities. Anyway, the model's drawback as compared to our KW+NE model is that it used only entity names but not all ontological features. Consequently, it cannot support queries searching for documents about entities of particular classes, name-class pairs, or identifiers.

## 6    Conclusion

We have presented and evaluated two extended VSMs that take into account different combinations of ontological features with keywords, namely the multi-vector model KW∪NE and the generalized term model KW+NE.  These two new models yield nearly the same performance, in terms of the precision, recall, and MAP measures, and are better than both the purely keyword-based model and the purely NE-based one. Our consideration of entity name aliases and class subsumption is logically sound and empirically verified. We have also taken into account and mapped interrogative words in queries to named entity classes, as in the proposed KW+NE+Wh model. That is intuitively justified and its advantage proved by the experimental results.

For its uniformity and simplicity, we propose the generalized term model for text retrieval. Meanwhile, the multi-vector model is useful for document clustering with respect to various ontological features. These are the first basic models that formally accommodate all entity names, classes, joint names and classes, and identifiers. Within the scope of this paper, we have not considered similarity and relatedness of generalized terms of keywords and named entities. This is currently under our investigation expected to further increase the overall performance of the proposed models.